\documentclass[prl,showpacs,preprintnumbers,tightenlines,twocolumn]{revtex4}
\usepackage[pdftex]{graphicx}
\usepackage{dcolumn}
\usepackage{bm}
\usepackage{epsfig}
\usepackage{stmaryrd}
\usepackage{dsfont}
\usepackage{bbm}
\usepackage{amsmath}
\usepackage{amsfonts}
\usepackage{amssymb}
\usepackage{amscd}
\usepackage{amstext}
\usepackage{float}
\usepackage{rotating,standalone}
\usepackage{color}
\usepackage{tikz}

\usetikzlibrary{calc}
\usetikzlibrary{positioning}
\usetikzlibrary{decorations.pathmorphing,patterns}
\usetikzlibrary{backgrounds}
\usetikzlibrary{matrix}

\newcommand{\vc}[1]{{\mathbf{#1}}}
\newcommand{\ssection}[1]{{\noi  \it #1:}}

\newcommand{\bra}[1]{\langle\,{#1}\, |}
\newcommand{\ket}[1]{|\,{#1}\,\rangle}
\newcommand{\braket}[2]{\mbox{$\langle\,{#1}\, | \,{#2}\,\rangle$}}

\newcommand{\tabvspace}{\Big.}

\newcommand{\sub}[2]{{#1}_{\mbox{\!\! \scriptsize #2}}}

\newcommand{\bv}[1]{\mathbf{ #1 }}

\def\noi{\noindent}
\def\beq{\begin{equation}}
\def\eeq{\end{equation}}

\def\CR{\nonumber\\[0.15cm]}

\def\figurewidth{0.99}
\newcommand{\rref}[1]{Ref.~\cite{#1}}
\newcommand{\fref}[1]{FIG.~\ref{#1}}

\newcommand{\frefp}[2]{FIG.~\ref{#1}#2}

\newcommand{\eref}[1]{Eq.~(\ref{#1})}

\newcommand{\cref}[1]{chapter~\ref{#1}}

\newcommand{\Cref}[1]{Chapter~\ref{#1}}

\newcommand{\bref}[1]{(\ref{#1})}

\newcommand{\op}[1]{\ensuremath{\hat{#1}}}
\newcommand{\im}{\ensuremath{\mathrm{i}}}

\newcommand{\sew}[1]{#1}
\newcommand{\karlo}[1]{#1}

\newcommand{\scensymm}{c}
\newcommand{\scenrep}{a}
\newcommand{\scenmid}{b}

\newcommand{\atomd}{2}
\newcommand{\atome}{3}
\newcommand{\atomf}{4}
\newcommand{\atomg}{5}
\newcommand{\atomh}{6}

\begin{document}
\title{Switching exciton pulses through conical intersections}
\author{K.~Leonhardt, S.~W\"uster and J.~M.~Rost }
\affiliation{Max Planck Institute for the Physics of Complex Systems, N\"othnitzer Strasse 38, 01187 Dresden, Germany}
\email{karlo@pks.mpg.de}
\begin{abstract}
Exciton pulses  transport excitation and entanglement adiabatically through Rydberg aggregates,
 assemblies of highly excited light atoms,  which are set into directed motion by resonant dipole-dipole interaction. Here, we demonstrate the coherent splitting of such pulses as well as the spatial segregation of electronic excitation and atomic motion. Both mechanisms exploit local non-adiabatic effects at a conical intersection, turning them from a decoherence source into an asset. The intersection provides a sensitive knob controlling the propagation direction and coherence properties of exciton pulses. The fundamental ideas discussed here have general implications for excitons on a dynamic network.
\end{abstract}
\maketitle

\ssection{Introduction}
%
Frenckel Excitons~\cite{frenkel_exciton}, in which excitation energy of an interacting quantum system is coherently shared among several  constituents, are a fundamental ingredient of photosynthetic light harvesting~\cite{book:maykuehn}. Recently, they have  become accessible in ultracold Rydberg gases, due to strong long-range dipole-dipole interactions \cite{park:dipdipbroadening,park:dipdipionization,li_gallagher:dipdipexcit,westermann:transfer,muelken:excitontransfer,bettelli:emerglatt,guenter:observingtransp,ravets:foersterdipdip,barredo:threespinchain}  and large lifetimes of atomic Rydberg states~\cite{book:gallagher,Beterov-Bezuglov-IonizationofRydberg-2009}. Assemblies  of several regularly placed and Rydberg excited cold atoms -- flexible Rydberg aggregates --  provide new concepts such as adiabatic guiding of an exciton through atomic chains as an exciton pulse~\cite{cenap:motion,wuester:cradle,moebius:cradle}. Such a pulse is initiated by a displacement of one atom in the regular chain.  The displacement simultaneously localizes the exciton on the atom pair with the smallest separation and initiates the pulse, i.e., the motion of the exciton. Subsequent binary collisions propagate the exciton pulse and the associated entanglement through the chain with very high fidelity. Without atomic motion
 this would require careful tuning of interactions \cite{asadian:motion,alex:heart,saikin:excitonreview,kuehn:excitonreview}.

That this propagation along a one-dimensional (1D) chain of atoms preserves the exciton with high fidelity \cite{wuester:cradle} is remarkable, since the transport appears quite fragile requiring a lossless locking in of electronic excitation transfer and atomic motion. However exciton transport often occurs in higher dimensional systems, where it is a priori unclear if we can also guide and control the exciton pulse.
Already a two-dimensional setup of atoms gives rise to conical intersections (CIs) \cite{tucker73_geometry_ci,mead79_bo_with_ci,wuester:CI} which might compromise adiabaticity known to be an important prerequisite of exciton transport. On the other hand CIs can play constructive roles in photochemical processes~\cite{domke:yarkony:koeppel:CIs}, hence might be similarly useful for atomic aggregates.

\begin{figure}[htb]
\centering
\epsfig{file=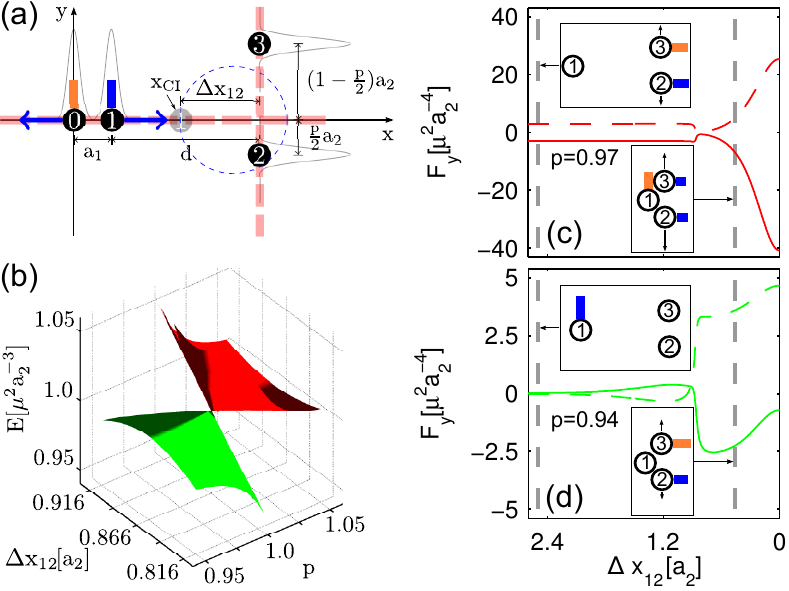,width= 0.99\columnwidth} 
\caption{(color online) (a) Orthogonal atom chains with one Rydberg dimer each.
 Atoms 0 and 1 initially share an excitation, due to which atom 1 reaches the conical intersection at $\sub{x}{CI}$. The origin of the coordinate system is set to the mean initial position of atom~0. (b)~The repulsive energy surface $U_{\mathrm{rep}}$ (red) and middle surface $U_{\mathrm{mid}}$ (green) of the trimer sub-unit (atom~1, 2 and 3) near the CI.
(c,d) Forces on atom~2 (solid lines) and atom~3 (dashed lines), for the repulsive surface (red,c) and middle surface (green,d). The insets show atomic positions and the excitation distribution ($d_n$, see text) of exciton states and forces for the indicated values of~$\Delta x_{12}$, which denotes the distance between atom~1 and the vertical chain.  The parameter $p$ controls the degree of symmetry of the trimer, where $p=1$ corresponds to an isosceles trimer configuration.
\label{geometry}}
\end{figure}
In the following, we will show how an exciton pulse can be coherently split through a CI that arises between two excitonic Born-Oppenheimer (BO) surfaces of the system. The junction between two atomic chains that gives rise to the conical intersection can be functionalized in two ways,  as a beam-splitter or a switch, sending the pulse split in both directions on the second chain or in only the one preselected. The surfaces involved in the CI serve as output modes of the beam-splitter. We will explicitly demonstrate that the junction constitutes a sensitive point where essential characteristics of the exciton pulse propagation can be controlled through small shifts in external trapping parameters.
Our results concern exciton transport on any network whose constituents move, such as (artificial) light-harvesting devices \cite{saikin:excitonreview}.

\ssection{T-shaped aggregates}
%
The junction is created  with two chains of Rydberg atoms in a T-shape configuration (\fref{geometry}a), with the required one-dimensional confinement generated optically \cite{li:lightatomentangle,rick:Rydberglattice}. 
Specifically, we will use  $2N$ Rydberg atoms with mass $M = 11000$ a.u.~and principal quantum number $\nu=44$. We assume that $N$ of these atoms are constrained on the $x$-axis, and the other $N$ on the $y$-axis, such that all atoms can only move freely in one dimension.  We start with one Rydberg atom  in an angular momentum $p$-state, while the rest are in $s$-states. The electronic wavefunction $\ket{\psi_{\rm{el}}(\vc{R})}$ of the whole system can be expanded in the single excitation basis 
$
 \ket{\psi_{\rm{el}}(\vc{R})}=\sum_{n=1}^{2N}c_n(t)\ket{\pi_n},
$
where $\ket{\pi_n}=\ket{s\dots p\dots s}$ is the state with the $n$-th atom in the $p$-state~\cite{cenap:motion,wuester:cradle,moebius:cradle} and $\vc{R}~=~(\vc{R}_1, \dots, \vc{R}_{2N})^{\rm{T}}$ groups all atomic coordinates $\vc{R}_n$.
The system is ruled by the Hamiltonian
\begin{align}
\sew{\op{H} =-\frac{\hbar^2}{2M}\nabla^2_{\vc{R}} +  \sum_{m\neq n=1}^{2N}\left( \op V_{\rm{dd}}(R_{mn})+\op{V}_{\rm{VDW}}(R_{mn})\right)\,,}
\label{eq:fullham}
\end{align}
 where $R_{mn} = |\vc R_m- \vc R_n|$ \karlo{is} the distance between atoms $m$ and $n$.

The long-range Rydberg-Rydberg interactions~\cite{book:gallagher} are described with two explicit contributions $\karlo{\hat{V}}_\mathrm{dd}$ and $\karlo{\hat{V}}_\mathrm{VDW}$, a resonant dipole-dipole and a van der Waals term, respectively.
$\karlo{\hat{V}}_\mathrm{dd}$ couples excitations $\ket{\pi_n}$ on different atoms $n$ through 
\begin{align}
\sew{ \op{V}_{\rm{dd}}(R_{mn})=-\frac{\mu^2}{R_{mn}^3}\ket{\pi_m}\bra{\pi_n}\,,}
 \label{eq:elechamiltonian}
\end{align}
where $\mu=d_{\nu,1;\nu,0}/\sqrt{6}$ is the scaled radial matrix element. The non-resonant van der Waals (VDW) interaction 
\begin{align}
 \op{V}_{\rm{VDW}}(R_{mn})=-\frac{C_6 }{2R_{mn}^6}\mathbb{I}\,,
 \label{eq:VdWhamiltonian}
\end{align}
ensures for $C_6<0$ repulsive behavior at very short distances regardless of the electronic state. Therefore,
 $\mathbb{I}$ denotes a unit matrix in the electronic space.  \sew{We sketch in \cite{sup:info} how this simple model of interactions arises from the full molecular physics of interacting Rydberg atoms~\cite{haroche:li_finesplitting} using a magnetic field and selected total angular momentum states.}

As previously shown \cite{cenap:motion,wuester:cradle,wuester:CI,moebius:cradle}, the joint motional and quantum state dynamics can be well understood 
from the eigenstates $\ket{\varphi_k(\vc{R})}$ of the electronic Hamiltonian $\sub{\hat{H}}{el}(\vc{R})=\sum_{m\sew{\neq} n}[\op V_{\rm{dd}}+\op{V}_{\rm{VDW}}]$.
These eigenstates and  the corresponding eigenenergies $U_k(\vc{R})$ depend parametrically on $\vc R$ and are referred to as Frenkel excitons~\cite{frenkel_exciton} and Born-Oppenheimer surfaces (BO surfaces), respectively.
The total  wavefunction including atomic motion can be written as $\ket{\Psi(\bv{R})}= \sum_n \phi_n(\vc{R}) \ket{\varphi_n(\vc{R})}$. To solve the coupled electronic and motional dynamics, we employ Tully's fewest switching algorithm \cite{tully:hopping2,tully:hopping:veloadjust,barbatti:review_tully,truhlar:tully_CI_acc,sup:info}, a quantum-classical method that is well established for our type of problem \cite{wuester:cradle,moebius:cradle,moebius:bobbels}.

\ssection{Two perpendicular dimers}
To realize a T-shape  chain, a minimum of two dimers is required, see \fref{geometry}a. The atoms have a Gaussian distribution about their initial location $\vc R_0$ along their chain with width $\sigma=0.5\ \mathrm{\mu m}$ as sketched. Transverse to the chain we assume perfect localization.
The bars in \frefp{geometry}{a} visualize the excitation amplitude of the exciton on the
repulsive BO surface  $\ket{\varphi_\mathrm{rep}(\vc{R}_0)}$. For each atom the length of the bar shows $d_n=\braket{\pi_n}{\varphi_\mathrm{rep}(\vc{R}_0)}$, \karlo{orange} for positive and blue for negative values. As one can see, initially the single p-excitation in the system is shared among atom~0 and 1. On the BO-surface $k$, the force on atom $n$ is given by $\bv{F}_{nk}=-\nabla_{\bv{R}_n} U_k(\vc{R})$.
Due to the initial repulsive force $\sub{F}{n,rep}$ (blue arrows) atom~1 moves and eventually reaches the position  $x_\mathrm{CI}$, where the  atoms~1-3 form a planar trimer. The two highest BO surfaces of this trimer conically intersect when the three atoms form an equilateral triangle, as shown in \frefp{geometry}{b} and studied in detail in \cite{wuester:CI}. In the following we will call these surfaces the repulsive (red) and middle (green) surface, respectively.

\begin{figure}[htb]
\centering
\epsfig{file=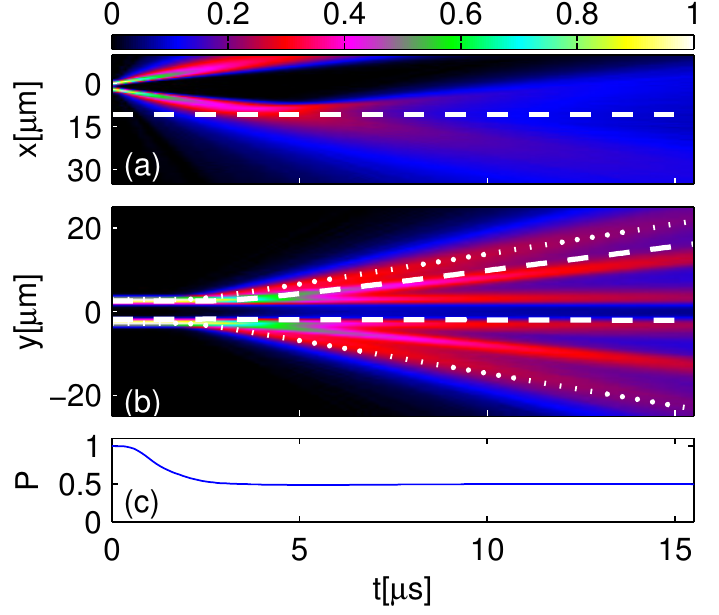,width= \figurewidth\columnwidth} 
\caption{(color online) Normalized total atomic density for atomic motion on two BO surfaces; overlayed are selected trajectories of the quantum classical method.  (a)~horizontal density $n(x,t)$ (see \cite{sup:info}) of atom~0 and~1. The dashed white line marks the x-position of the vertical chain. (b)~vertical density $n(y,t)$ of atom~2 and~3. In (a,b) we actually plot $\sqrt{n(x,t)}$, $\sqrt{n(y,t)}$. (c) Purity of reduced electronic state, see \cite{sup:info}.
Our calculations are for lithium atoms excited to principal quantum number
  $\nu=44$ with a transition dipole moment of $\mu=1000$~a.u.. The parameters
  of the initial atomic configuration were  $a_1=2.16\ \rm{\mu m}$,
  $a_2=5.25\ \rm{\mu m}$ and $d=8.5\ \rm{\mu m}$. To sample the nuclear
  wavefunction, we have propagated $10^5$ trajectories~\cite{xmds} with a standard deviation of
  $\sigma_x=0.5\ \rm{\mu m}$ for the atomic positions. Here, we have used $C_6=0$ \sew{for simplicity}.
\label{splitting}
}
\end{figure}

\ssection{Exciton splitting}
 Initialized on the repulsive surface of the global (double dimer) system, the exciton pulse is transferred to the vertical chain via the conical intersection onto these two electronic surfaces --
 the repulsive and the middle one -- dependent on the position of atom 1
 relative to atom 2 and 3 in the y-direction when it enters the trimer configuration (see parameter $p$ in \frefp{geometry}a).
 Viewed from the perspective of the trimer subsystem only, the exciton pulse enters on the \emph{middle} surface, where the excitation amplitude (blue bar) matches the initial excitation distribution, see insets of \frefp{geometry}{c,d}.
   If atom 1 arrives right in the middle between atoms 2 and 3, the atomic trimer passes through the degenerate point of the CI leading to significant transfer of exciton amplitude to the repulsive surface (\frefp{geometry}{c}). If this is not the case, an asymmetric trimer configuration is realized for which non-adiabatic transitions due to the CI are much weaker and the system remains on the middle trimer surface leading to the situation of \frefp{geometry}{d} with quite different forces  on atom 2 and 3. This has profound consequences on the atomic motion: Amplitude on the repulsive surface leads to a symmetric repulsion of atoms 2 and 3 of the vertical chain, creating the outer pulses in the density shown in \frefp{splitting}{b}. A representative quantum-classical trajectory is shown as white dotted line, with $p=0.98$.
On the repulsive surface atom 1 is often reflected off the vertical chain as visible in \fref{splitting}{a}.
On the other hand, amplitude on the middle surface has the effect of a very asymmetric atomic motion in y, with that atom on the y-axis remaining almost at rest which has initially the smaller distance to the location of the dimer on the x-axis. This type of motion is responsible for the inner and central features in \frefp{splitting}{b}, with a representative trajectory shown white dashed with $p=0.82$. One can also recognize the variant of the motion, where the other vertical atom remains at rest. The middle surface is mainly responsible for atom 1 freely passing the vertical chain in \fref{splitting}{a}.
  
 Since the  nuclear wave packet of the exciton pulse will have a distribution of positions of atoms 1, 2 and 3, 
 there will be in general a splitting of the exciton when it has passed the conical intersection
 with the electronic excitation  propagating on the repulsive as well as the middle surface. In fact,
 about 50\% of the initial amplitude has been transferred from the middle to the repulsive surface after 
 $3\ {\mu}$s under the initial conditions for our exciton pulse leading to the dynamics of \fref{splitting}.

Hence, not only does the exciton  split into two parts traveling with the atoms in opposite directions in the y-chain,
we have a further coherent splitting of electronic excitation into the middle and repulsive electronic surface:   The total initial wave function  was $\ket{\sub{\Psi}{ini}(\bv{R})}= \sub{\phi}{0}(\vc{R}) \ket{\sub{\varphi}{rep}(\vc{R})}$, where $\sub{\phi}{0}(\vc{R})$ describes the initial, harmonically trapped, spatial ground state. After the evolution shown in \fref{splitting}, the wave function reads $\ket{\sub{\Psi}{fin}(\bv{R})}= \sub{\phi}{rep}(\vc{R}) \ket{\sub{\varphi}{rep}(\vc{R})}+ \sub{\phi}{mid}(\vc{R}) \ket{\sub{\varphi}{mid}(\vc{R})}$.

In this final state $\ket{\sub{\Psi}{fin}(\bv{R})}$ the atomic configuration and the electronic state are entangled, which can be 
 quantified  through the purity of the reduced electronic density matrix. It is obtained by averaging over the atomic position  as described in \cite{sup:info,wuester:cradle,moebius:cradle}. The purity drops from one to $1/2$ when the exciton is split, as shown in \fref{splitting}d, reflecting a transition from a pure to a mixed state. For the total (pure) system state, this  implies a transition from a separable to an entangled state.

\ssection{Exciton switch}
 The minimal T-shape system consisting of two dimers  discussed so far primarily serves the purpose to 
elucidate the central element for exciton pulse control, namely the junction between perpendicular atomic chains.
Ultimately, we would like to interface the two dimers 
 in \frefp{geometry}{a}  with longer atomic chains that can support exciton pulses as described in \cite{wuester:cradle}. Such a pulse travels to the junction to become coherently split as just described, with the resulting exciton pulses on the vertical chain depending on how the conical intersection of the trimer at the junction was passed.
Since the relative strength of the exciton pulse on the middle and repulsive surface of the trimer  depends on the atomic positions and momenta near the conical intersection, we can control the exciton pulse propagation on the vertical chain, for example via the position of the horizontal chain relative to the vertical one. We demonstrate this effect with {3~atoms on the horizontal and 4 atoms on the vertical chain}, allowing for a vertical offset $\Delta y$ of the horizontal chain from the center of the vertical chain, and a variable separation $a_2$ of the two central atoms in the second chain, see sketch in \fref{entanglementswitching}.
\begin{figure}[htb]
\centering
\epsfig{file=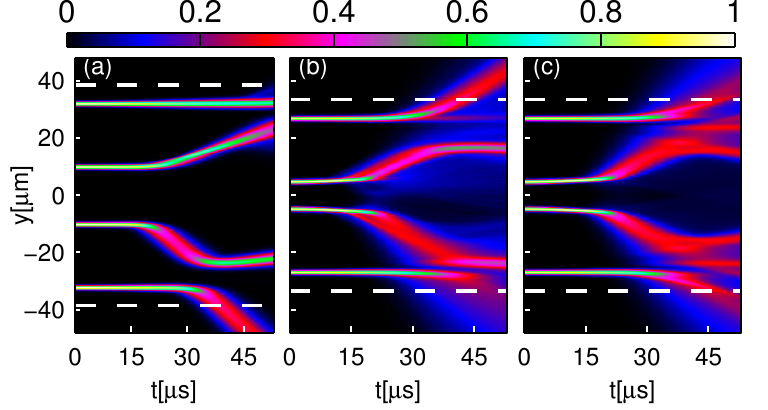,width= \figurewidth\columnwidth} 
\caption{(color online) Use of the trimer subunit with conical intersection as an exciton switch. Depending on the geometry, we can realize three qualitatively different scenarios: (\scenrep) asymmetric, repulsive surface; (\scenmid) asymmetric, middle surface; (\scensymm) symmetric, repulsive and middle surface. Each plot shows $\sqrt{n(y,t)}$ as in \fref{splitting}. When atoms reach the white horizontal lines ($\sub{r}{coll}${, $6.6\mu$m beyond the initial $y$-position of atoms \atome, \atomh}), we extract the level of bi-partite entanglement, listed in \fref{entanglementswitching}. 
The parameters different to those of \fref{splitting} are $\nu=80$ (hence
  $\mu=3374$~a.u.), $a_1={6}{\mu}$m, $d={22}{\mu}$m and $C_6\approx{-7.6\cdot 10^{20}}$~a.u.~as derived in \cite{sup:info}. Also see \cite{sup:info} for videos of representative single trajectories.
\label{switch}
}
\end{figure}

 With small variations of the two parameters $\Delta y$, $a_2$, qualitatively very different scenarios can be realized  as illustrated with \fref{switch} and \fref{entanglementswitching}.
 
 In scenario (\scenrep), the spacing $a_2$ { and the shift $\Delta y$ } {are} so large that the trimer subunit discussed before does not form \cite{footnote:notrimer}. 
Since atom $\atomd$ approaches  $\atomf$ closest (see \fref{entanglementswitching}), the exciton-pulse travels in the downwards direction. To switch it upwards we would use $\Delta y\rightarrow-\Delta y$.
We characterize the relevant entanglement transport using $\bar{E}_{ij}$, the bi-partite entanglement \cite{hill:wootters:qbits,wootters:mixed,sup:info,wuester:cradle,moebius:cradle} during the last collision of the two terminal atoms $i,j$  on the vertical chain, i.e., $(i,j)=(\atomg,\atomh)$ upwards and $(i,j) =(\atome,\atomf)$ downwards. We know the last collision is in progress, whenever atom $\atome$ (atom $\atomh$) reaches  position $\sub{r}{coll}$, indicated in \fref{switch} by horizontal white lines for the downwards (upwards) direction. Our results in \fref{entanglementswitching} reveal that 
{pulse propagation is linked with high fidelity entanglement transport,}
demonstrating successful  control of the direction of  exciton-pulse propagation without loosing  coherence.

In scenario (\scenmid) we  segregate mechanical and electronic degrees of freedom of the exciton-pulse, by choosing  $a_2$ small enough such that a trimer subunit forms at the junction. Since an offset $\Delta y$ is kept, the nuclear wave packet, however, misses the conical intersection and remains on the middle trimer surface.
Importantly, the middle trimer surface does not connect to a global surface allowing coherent exciton pulse transport: At the first collision within the vertical chain (between atoms $\atome$ and  $\atomf$ in \frefp{switch}{\scenmid}), part of the excitation evades those atoms and delocalizes on the remnant {upper} chain. Momentum is henceforth transported {downwards} by van der Waals collisions only such that the original exciton pulse with entangled atom and electron dynamics has been ripped apart. {For this momentum transport without excitation transport, the inclusion of VDW interactions is crucial. They further cause the atom 
on the y-axis closest to the x-axis to carry most acceleration, in contrast to \fref{splitting}.}

Finally, in scenario (\scensymm) $\Delta y=0$ and the wave packet fully traverses the conical intersection at the junction. Here, the trimer subunit  operates as described in the first part of the article. The wave packet is split onto both, the repulsive and middle trimer surface. As discussed for scenario (\scenmid), the middle trimer surface does not give rise to exciton-pulse propagation. On the repulsive surface, one gets  symmetric (up-down) propagation of two pulses as expected. However, the entanglement transport in both directions is  much weaker than in scenario (\scenrep) which is due to the fact that
 the atoms still share only \emph{a single} p-excitation. Subsequent non-adiabatic effects  allow a strong coherent pulse only in a single direction.  Even \emph{within} this symmetric scenario, the relative importance of the middle and repulsive surface can be tuned via the effective size of the conical intersection \cite{teller1937crossing}, determined by atomic velocities and separation (energy splittings).

\begin{figure}[htb]
\begin{center}
\begin{tabular}{cc} 
\centering
\epsfig{file=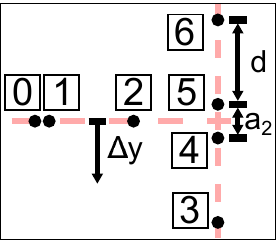,width= 0.33\columnwidth} 
& 
\centering
\begin{tabular}{|c|ccc|}
 \vspace{-2.75cm}\\
\cline{1-4}
 \tabvspace scenario      &(\scenrep)      & (\scenmid)  & (\scensymm)   \\
\cline{1-4}
$a_2$    & ${20}\mu$m & ${9.5}\mu$m & ${9.5}\mu$m \\
${\Delta}y$  & ${1.5}\mu$m & ${1.5}\mu$m & 0 \\
\cline{1-4}
up, $\tabvspace \bar{E}_{56}$    & $0\%$ & ${60}\%$ & ${24}\%$ \\
down, $\tabvspace \bar{E}_{34}$  & ${97}\%$ & ${7}\%$ & ${24} \%$ \\
\cline{1-4}
\end{tabular}
\end{tabular}
\end{center}
\caption{Geometry and entanglement switching for the three scenarios of \fref{switch}. Atom numbering and control parameters $a_2$, ${\Delta}y$ are defined in the sketch on the left. The entanglement measure $\bar{E}_{ij}$ is defined in \cite{sup:info}.
\label{entanglementswitching}}
\end{figure}
%

\ssection{Conclusions}
%
We have shown how an exciton pulse can be coherently split through non-adiabatic dynamics at a conical intersection in a flexible Rydberg aggregate.
Our results turn a junction between two Rydberg atom chains into a switch. 
 The switch can control if and how exciton pulses continue to propagate in the system.
Similar physics may be of interest for research on artificial light harvesting systems \cite{mcconell:artificialLHCreview}, where exciton transport and control is quintessential for energy efficiency. The atomic junction introduced  here also provides a tool to directly examine the many-body dynamics near conical intersections in the laboratory.

The exciton splitting predicted could  be experimentally monitored using high resolution Rydberg atom detection schemes~\cite{schwarzkopf:correlations,olmos:amplification,guenter:EIT} which are in addition state selective.  Applied to our system, they  allow a direct visualization of many-body wave packet dynamics near a conical intersection. The essential modular subunit of an atomic trimer exhibiting a CI can be envisaged as a building block for networks of exciton carrying atomic chains or a device for controlling the energy flow in molecular aggregates.

\acknowledgments
We gratefully acknowledge fruitful discussions with Alexander Eisfeld and Sebastian M{\"o}bius, as well as financial support by the Marie Curie Initial Training Network COHERENCE.

\ssection{\bf{Supplemental material: Switching exciton pulses through conical intersections}}

{\it This supplemental material provides additional details regarding the employed quantum-classical algorithm, the Rydberg trimer subunit, our purity and entanglement measure, extraction of total atomic densities and the realization of isotropic dipole-dipole interactions.}

\ssection{Propagation}
\label{propagation}
For larger number of atoms, solving the time dependent Schr\"odinger equation for our problem is not feasible in a reasonable time.
However a quantum-classical propagation method, Tully's fewest switching algorithm \cite{tully:hopping2,tully:hopping:veloadjust,barbatti:review_tully}, gives results in good agreement with the full propagation of the Schr\"odinger equation \cite{wuester:cradle,moebius:cradle,moebius:bobbels}.
In Tully's fewest switching algorithm the positions of the atoms $\vc{R}$ are treated classically while their electronic state is described quantum mechanically. To retain further quantum properties two features are added.
First, the atoms are randomly placed according to the Wigner distribution of the initial nuclear wavefunction and also receive a corresponding random initial velocity.
In the end of the simulation, all observables have to be averaged over the whole set of realizations.
Second, non-adiabatic processes are added as follows:
During the propagation of a single realization, the mechanical potential felt by the atoms corresponds to a single eigenenergy of the electronic Hamiltonian. During adiabatic processes the system remains on a single energy surface during the propagation. Tully's algorithm allows for jumps to other energy surfaces during the propagation. The probability for a jump from surface $n$ to surface $m$, is proportional to the non-adiabatic coupling vector
\begin{equation}
 \vc{d}_{mn}(\vc{R})=\bra{\varphi_m(\vc{R})}\nabla_{\vc{R}}\ket{\varphi_n(\vc{R})}.
\end{equation}
The sequence of propagation is as follows: The positions and velocities of the atoms are randomly determined. The electronic Hamiltonian is diagonalized and we use one eigenenergy $U_k$ of our choice as potential for the atoms. The atoms are propagated one time step via Newton's equation
\begin{equation}
 M\ddot{\vc{R}}=-\nabla_{\vc{R}}U_k(\vc{R}).
\label{eq:newton_equation}
\end{equation}
The new positions lead to new eigenstates and -energies and to new diabatic and adiabatic coefficients. We propagate the diabatic coefficients via
\begin{equation}
 \im \hbar \dot{c}_k(t)=-\mu^2\sum_{l\neq k}^{2N}\frac{c_l(t)}{|\vc{R}_k-\vc{R}_l|^3}
\label{eq:propagation_diabatic_states}
\end{equation}
To close the loop, the nuclei will be propagated via \eqref{eq:newton_equation} again.\newline
We imagine the atoms were confined in individual harmonic traps, before these are released to let all atoms move. This motivates Gaussian probability distributions of the atomic positions and momenta. We label the standard deviation in position of atoms on chain $i$ by $\sigma_{x_{i}}$. The velocity probability distribution then has a standard deviation $\sigma_{v_i}=\hbar/(M \sigma_{x_i})$. 

\ssection{Validation}
\label{validation}
\begin{figure}[htb]
\centering
\epsfig{file={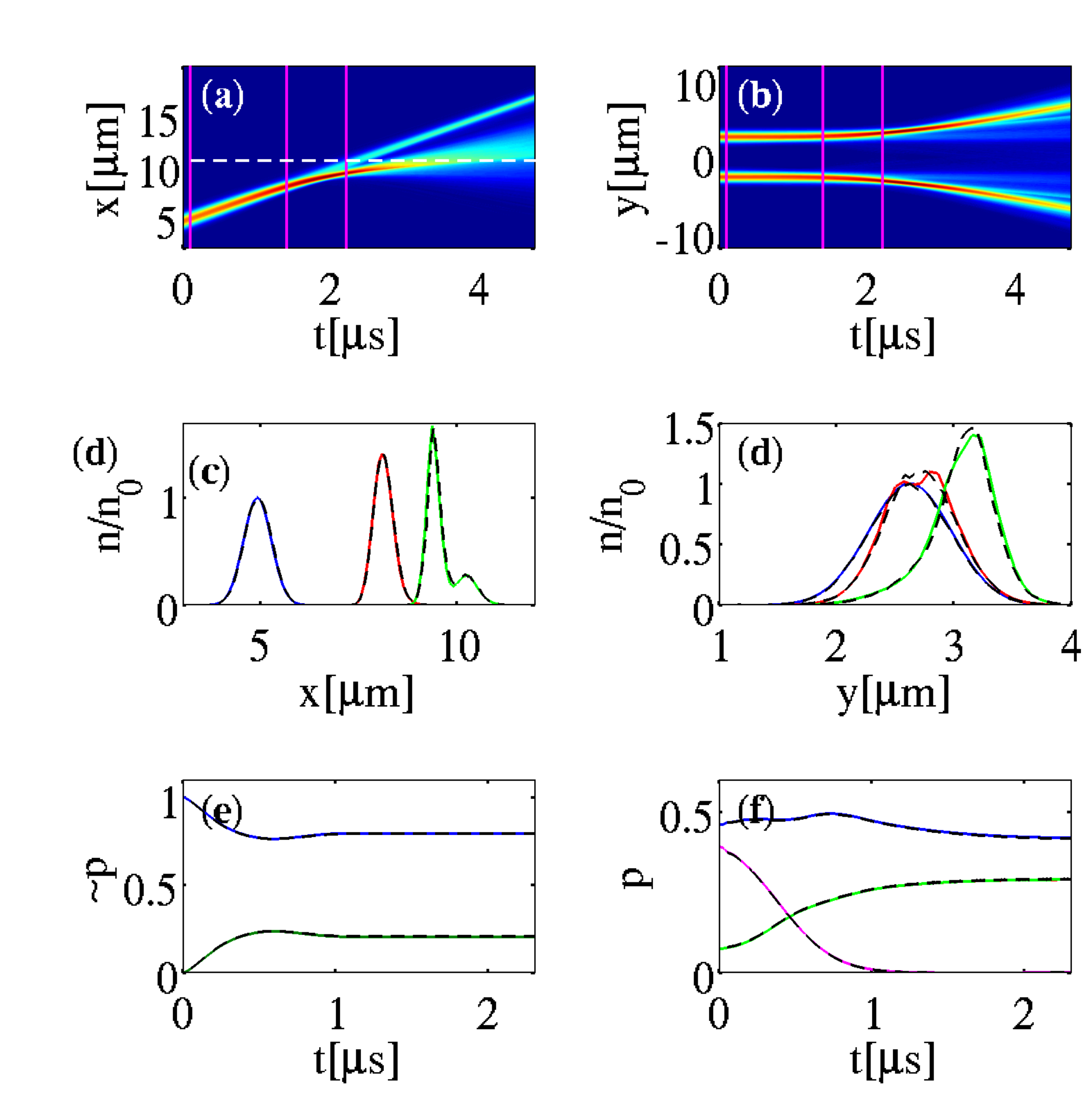},width=0.99\columnwidth} 
\caption{\label{tully_versus_QM}
Comparison of Tully's surface hopping with full quantum mechanical calculations, for exciton switch similar to Fig.~2 of the main article. (a) Horizontal density $\sqrt{n(x,t)}$ of atom~1 from Tully's method. Vertical lines indicate snap shots shown in panel (c). The dashed white line marks the location of the vertical chain. (c) Horizontal density $n(x,t_k)$ of atom~1 from Schr{\"o}dinger equation (solid) and Tully's surface hopping (black dashed). (blue) $t_1=0$, (red) $t_2=1.4 \mu$s, (green) $t_3=2.2 \mu$s.
(b,d) The same for the density on the vertical chain $\sqrt{n(y,t)}$, $n(y,t_k)$ of atom~2 and 3. (e) Population on Born-Oppenheimer surfaces (color) quantum, (black-dashed) Tully. (f) Excitation probability on each atom, quantum: (magenta) atom 0, (blue) atom 1, (green) atoms 2 and 3, (black-dashed) Tully.
}
\end{figure}
Tully's fewest switching algorithm discussed above has already been benchmarked successfully for exciton dynamics on Rydberg chains in \cite{wuester:cradle,moebius:cradle,moebius:bobbels}. For the exciton switch discussed here, an essential new ingredient is the conical intersection resulting in strong non-adiabatic effects. We show here that these are captured by Tully's method very well, by comparison with full quantum-mechanical calculations. To make the strongest connection with the present work, we study a scenario close to the exciton switch shown in Fig.~2.

Modifications that were necessary to keep quantum simulations tractable are a freezing of the motion of atom 0, and the removal of its position uncertainty. Further the initial acceleration period is removed, instead atom 1 is shifted by $\sub{x}{shift} = 2.5\mu$m towards positive $x$ and given an initial mean velocity $\sub{v}{shift} = 2.5$m/s. This is about half of what would correspond more closely to Fig.~2, however larger velocities would require too fine numerical grids. As described in [19], we then solve a multi-component Schr{\"o}dinger equation in the electronic $\ket{\pi_n}$ basis.

As seen in \fref{tully_versus_QM}, both exciton- and spatial dynamics are captured satisfactorily by the quantum-classical method. The performance of Tully's method has been intensively studied in the context of quantum chemistry (e.g.~[27]). For the dynamics of exciton transport on moving (flexible) Rydberg assemblies studied here, the present comparison and that of \rref{wuester:cradle} shows excellent agreement. A distinguishing feature of our systems is that spatial interferences on any BO surface typically do not occur, thus spatial coherence information that is not included in Tully's method is not required.

\ssection{The trimer}
\label{the_trimer}
``Trimer'' refers to an assembly of three Rydberg atoms. Since main features of the systems we studied can be understood by considering only three atoms, we provide here full details in addition to the features discussed in the main text.
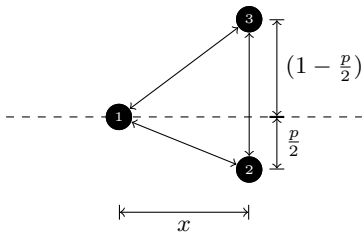
\begin{figure}[htb]
\centering
\begin{tikzpicture}
\tikzset{atoms/.style={inner sep=0pt,minimum size={#1},black,fill,circle}};

\node[atoms={0.35cm},label={[white]center:{\tiny{1}}}] (a1) at (0,0) {};
\node[atoms={0.35cm},label={[white]center:{\tiny{2}}}] (a2) at ($ (1.732,-0.7) $) {};
\node[atoms={0.35cm},label={[white]center:{\tiny{3}}}] (a3) at ($ (1.732,1.3) $) {};

\draw[dashed] ($ (a1)- (1.5,0)$) -- ($ (a1)+ (3.23,0)$) ;

\node[atoms={0.35cm},label={[white]center:{\tiny{1}}}] (a1) at (0,0) {};
\node[atoms={0.35cm},label={[white]center:{\tiny{2}}}] (a2) at ($ (1.732,-0.7) $) {};
\node[atoms={0.35cm},label={[white]center:{\tiny{3}}}] (a3) at ($ (1.732,1.3) $) {};

\draw[|<->|] ($(2.1,-0.7)$) -- (2.1,0)  node[right,midway] {$\frac{p}{2}$};
\draw[|<->|] (2.1,0) -- ($ (2.1,1.3) $) node[right,midway] {$(1-\frac{p}{2})$};
\draw[|<->|] ($(0,-1.268 )$) -- ($ (1.732,-1.268 )$) node[below,midway] {$x$};

\draw[<->] (a1) -- (a2) node[midway,sloped,below] {};
\draw[<->] (a1) -- (a3) node[midway,sloped,above] {};
\draw[<->] (a2) -- (a3) node[midway,sloped,above] {};
\end{tikzpicture}
\caption{\label{sketch_asymmetric_trimer}
Sketch of a trimer near a configuration with CI. Atom~1 is confined on a horizontal line and atom~2 and~3 on a vertical line. The parameter~$p$ adjusts the distance of atom~2 and~3 from the horizontal line and for $p\neq 1$ results in nonequilateral triangle configurations. All distances are expressed in units of $\mathrm{u}$, the distance between atom~2 and~3.
}
\end{figure}
The configurations of the trimer which are most relevant here are shown in \fref{sketch_asymmetric_trimer}. We call the overall length scale $\mathrm{u}$. 
The geometry of the trimer around the equilateral triangle configuration is described by the distance $x$ between atom~1 and the other two atoms and the parameter
\begin{equation}
p:=2\frac{|\langle\vc{R}_2,\vc{e}_y\rangle|}{|\vc{R}_2-\vc{R}_3|},
\end{equation}
which we call the asymmetry parameter, since it controls the degree of symmetry with respect to the isosceles triangle. The biggest and smallest eigenenergy are globally repulsive or attractive, respectively~\cite{cenap:motion,wuester:CI}. We label them $U_{\rm{rep}}$ and $U_{\rm{att}}$ and the corresponding eigenstates $\ket{\varphi_{\rm{rep}}}$ and $\ket{\varphi_{\rm{att}}}$. We call these repulsive and attractive surface and eigenstate, respectively. There is another eigenenergy energetically between them. We label it $U_{\rm{mid}}$ and the corresponding eigenstates $\ket{\varphi_{\rm{mid}}}$. We call it middle surface and eigenstate.

\begin{figure}[htb]
\centering
\epsfig{file={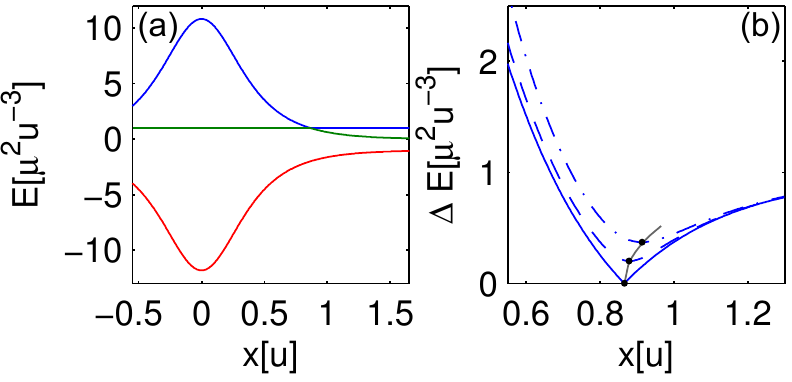},width=0.89\columnwidth} 
\caption{\label{fig:trimer_symmetric_asymmetric_energie_properties}
Eigenenergy spectra for the trimer.
(a)~Eigenenergies over horizontal distance x for the symmetric case $p=1$. The repulsive (blue line) and middle eigenenergy (green line) cross at x=$\sqrt{3}/2\ \mathrm{u}$.
(b)~Energy spacing between repulsive and middle eigenenergy for different asymmetry parameters, $p =1$ (solid), $p=0.88$ (dashed), $p=0.76$ (dashed dotted). The minimal energy spacing (black dots) is shifted to bigger $x$ for higher asymmetry, which is well described by the analytical result~\eref{eq:analyt_delta_E_min} (grey line).
}
\end{figure}

\paragraph{Symmetric case $p=1$.}
The middle and repulsive eigenenergies have the value $\mu^2\mathrm{u}^{-3}$, when they cross. This happens, when $x=x_{CI}:=\sqrt{3}/2\ \mathrm{u}$, i.e. at the equilateral triangle configuration. It is well known that this is a conical intersection~\cite{wuester:CI,tucker73_geometry_ci,mead79_bo_with_ci}. \fref{fig:trimer_symmetric_asymmetric_energie_properties}~(a) shows the eigenergies as a function of the horizontal distance $x$. The middle eigenenergy stays constant for $x < x_{CI}$ as it arises solely from the interaction energy of atom 2 and 3. When atom 1 is far away from the other two, the middle and attractive energies  are vanishing, whereas when the system realizes a linear trimer ($x=0$), the repulsive and attractive energy values are extremal.

\paragraph{Asymmetric case $p\neq 1$.}
There is no crossing of eigenvalues for $p\neq1$. 
\fref{fig:trimer_symmetric_asymmetric_energie_properties} (b) shows the energy separation between the repulsive and the middle state over the horizontal distance $x$ of the atoms for different asymmetry parameters $p$. With increasing asymmetry, the smallest energy splitting increases, as does the value of $x$ where the splitting is smallest.
From now on we call atomic configurations asymmetric, when they correspond to values of $p \ll 1 $ and symmetric, when $p \approx 1$.\newline

Using the parameters just defined, we can analyse the forces on the atoms for the two relevant BO-surfaces and find characteristically different behaviour as shown in Fig.~1 of the main article and discussed therein.

For the trimer, it is well known that where the atoms build an equilateral triangle, the energy surfaces exhibit a CI.
We now analyze the eigenenergies and eigenstates near the CI in order to understand the numerical results in~\fref{fig:trimer_symmetric_asymmetric_energie_properties}.
All different geometries of the trimer around the conical intersection can be described by the two parameters $\Delta x:=x-x_{CI}$ and $p$, illlustrated in~\fref{sketch_asymmetric_trimer}. We collect both in the vector $\vc{\chi}~:=~[\Delta x,p]^{\rm{T}}$.
The equilateral triangle configuration corresponds to  $\vc{\chi}_0~:=~[0,1]^{\rm{T}}$, with the degenerate eigenvalue $E^{(0)}_{CI}~=~\mu^2/\mathrm{u}^3$. The corresponding eigenstates can be $\ket{\varphi^{(0)}_{CI,1}}~=~\frac{1}{\sqrt{2}}\begin{bmatrix}-1 & 0 &1 \end{bmatrix}^{\rm{T}}$ and $\ket{\varphi^{(0)}_{CI,2}}~=~\frac{1}{\sqrt{6}}\begin{bmatrix}-1 & 2 &-1 \end{bmatrix}^{\rm{T}}$. 
The electronic Hamiltonian of the Configuration shown in \fref{sketch_asymmetric_trimer} is given by
\begin{equation}
\op{H}_{\rm{el}}(\vc{\chi}) = -\mu^2\begin{bmatrix}
0 			& s_1^{-3}(\vc{\chi}) 	& s_2^{-3}(\vc{\chi}) 	\\
s_1^{-3}(\vc{\chi}) 	& 0 			& s_3^{-3} 		\\
s_2^{-3}(\vc{\chi}) 			& s_3^{-3} 		& 0		\\                  
\end{bmatrix}.
\end{equation}
We use degenerate perturbation theory to estimate the energy gap near the CI. To do so we first Taylor expand the electronic Hamiltonian around the CI configuration up to second order in $\vc{\chi}$:
\begin{equation}
 \op{H}^{\rm{as}}_{\rm{el}}(\vc{\chi}) \approx \op{H}^{\rm{CI}}_{\rm{el}} + \op{H}^{\rm{PT}}_{\rm{el}}(\vc{\chi}),
\end{equation}
where $\op{H}^{\rm{CI}}_{\rm{el}}$ is the electronic Hamiltonian at the CI configuration and $\op{H}^{\rm{PT}}_{\rm{el}}(\vc{\chi})$ is the perturbation.
We define the perturbation matrix
\begin{equation}
S_{\rm{el}}(\vc{\chi})_{\alpha\beta}:=\bra{\varphi^{(0)}_{CI,\alpha}}\op{H}^{\rm{PT}}_{\rm{el}}(\vc{\chi})\ket{\varphi^{(0)}_{CI,\beta}}.
\end{equation}
The eigenenergies $E^{\rm(1)}_{1}(\vc{\chi}), E^{\rm(1)}_{2}(\vc{\chi})$ of $S_{\rm{el}}(\vc{\chi})$ are the first order corrections to the energy and lift the degeneracy. Thus the energy gap is given by $\Delta E^{\rm{as}}(\vc{\chi})~=~| E^{\rm(1)}_{1}(\vc{\chi}) - E^{\rm(1)}_{2}(\vc{\chi})|$ to first order. Consistently expanding this expression to second order around $\vc{\chi}_0$, we get
\begin{equation}
\dfrac{\Delta E^{\rm{as}}(\vc{\chi})}{\mu^2}\approx \sqrt{12\Delta x^2 + (1-p)^2\biggl(3-\frac{31\sqrt{3}\Delta x}{2} + \frac{1061\Delta x^2}{8}\biggr)}
\end{equation}
for $p\lesssim 1$. The asymmetry of the configuration is measured by $1-p$. For every small given asymmetry, there is a $\Delta x_{\rm{min}}$ where the energy gap becomes minimal:
\begin{align}
\begin{split}
 \Delta x_{\rm{min}} &\approx 1.12\cdot(1-p)^2\\
\Delta E^{\rm{as}}_{\rm{min}}(p)& \approx \sqrt{3}\cdot(1-p)
\label{eq:analyt_delta_E_min}
\end{split}
\end{align}
Thus the horizontal distance between atom 1 and the other atoms has to be bigger compared to the CI configuration, to achieve a minimal energy gap as evident in \fref{fig:trimer_symmetric_asymmetric_energie_properties}.

\ssection{Entanglement measure}
\label{entanglement}
As described in more detail, the quantum mechanical electronic density matrix $\hat{\sigma}=\sum_{n,m}\sigma_{nm} \ket{\pi_{n}}\bra{\pi_{m}}$ is represented by 
\begin{align}
\sigma_{nm}&=\overline{c_{n}^* c_{m}},
\label{red_el_Dm_ME}
\end{align}
in a quantum-classical framework, where $\overline{\cdots}$ denotes the trajectory average. The purity $P=\mbox{Tr}[\hat{\sigma}^2]$ quantifies to which extent the reduced electronic state is mixed ($P=0$) or pure ($P=1$).

We can further obtain a reduced density matrix for just two atoms
\begin{equation}
\hat{\beta}_{ab}={\mbox{Tr}}^{\{a,b\}}\big[\hat{\sigma}\big].
\label{beta}
\end{equation}
by performing the trace ${\mbox{Tr}}^{\{a,b\}}\big[\cdots\big]$ over the electronic states for all atoms other than $a$, $b$. For a single p-excitation in the system, this takes the form
\begin{eqnarray}
\hat{\beta}_{ab}=\left(
\begin{array}{cccc}
0 & 0 & 0 & 0 \\
0 & \sigma_{aa} &  \sigma_{ba} & 0 \\
0 &  \sigma_{ab}^* &  \sigma_{aa} & 0 \\
0 & 0 & 0 & \sum_{c\neq \{a,b\}} \sigma_{cc}  \\
\end{array}
\right).
\label{betafull}
\end{eqnarray}

The density matrix $\hat{\beta}_{ab}$ may describe mixed versions of entangled states, the entanglement of which is often quantified using $E_{ab}$, the ``entanglement of formation'' \cite{hill:wootters:qbits,wootters:mixed}. It is obtained through the concurrence $C_{ab}=2|\sigma_{ab}|$, with the further definitions
$H(x)=-[x \log_{2} x + (1-x) \log_{2}(1-x)]$ and 
${\cal E}(x)=H(1/2 + \sqrt{1-x^2}/2)$ as 
$E_{ab}={\cal E}(C_{ab})$.

Fig.~4 of the main article then shows the bi-partite entanglement of formation for the two last atoms on the vertical chain, in the respective direction as indicated.
\ssection{Extraction of total atomic density}
\label{atomic_density}
Fig.~2 and Fig.~3 show the atomic densities on the vertical and horizontal chain. In Tully's semiclassical method we propagate many individual trajectories with different atom positions, to sample the atomic wavefunction. To obtain total densities, we bin the coordinate of the atoms on the horizontal chain into a discrete grid for the x-axis and for atoms on the vertical chain into a discrete grid for the y-axis. This is averaged over all trajectories. By dividing through the number of atoms per chain $N$, we obtain the normalized total density for each chain. The formula for the x-axis density $n(x,t)$ reads:
\begin{align}
 n(x,t)&=\sum\limits_{k=1}^{N_{\mathrm{trajs}}}\sum_{j,m} \Theta\Bigl(\frac{\Delta}{2}-|x_m-x|\Bigr),
\CR
& \times \Theta\Bigl(\frac{\Delta}{2}-|R^{(k)}_{j,x}(t)-x_m|\Bigr) /NN_{\mathrm{trajs}},
\end{align}
where $\Theta(x)$ is the Heaviside function, the sum $\sum_j $ is over atoms on the horizontal chain only and the sum $\sum_m$ over all discrete bins on the x-axis.  
We used $R^{(k)}_{j,x}(t)$  for the x-coordinate of the $j$'th atom from the $k$'th trajectory and $x_m$ for the central bin coordinates. The binning grid spacing is $\Delta$.
We now have $\sum_{k=1}^{(R_{f}-R_{i})/\Delta} n(R_{i}+(k+1/2)\Delta,t)=1$, where $R_{i},R_{f}$ are the spatial boundaries of our binning.
The definition of the y-axis density $n(y,t)$ is analogous.

\ssection{Isotropic dipole-dipole interactions}
\label{isotrope_dip_dip_int}
In all simulations we used an electronic basis with a single p-excitation and assumed the dipole-dipole interaction to be isotropic, only dependent on the internuclear distance between two atoms, which we denote here with $R$. 
If spin-orbit interaction is neglected and the sign of the interaction irrelevant, this situation is achieved by choosing the quantization axis $\hat{z}$ perpendicular to our internuclear distance vectors and considering the magnetic quantum number $m=0$ manifold, which decouples from the others~\cite{moebius:cradle}.
 
For the simulations of Fig.~2 the principal quantum number was $\nu=44$, which yields a finestructure-splitting of $\Delta E_{\mathrm{FS}}=0.92$~MHz~\cite{haroche:li_finesplitting}. The characteristic strength of the dipole-dipole interaction $V_{dd}=d_{\mathrm{rad}}^2(44)/R^3~=~613$~MHz depends on the distance between the atoms and the radial matrix element $d_{\mathrm{rad}}(\nu):=d_{\nu,1;\nu,0}$  between $l=0$ and $l=1$ states. Here we used the initial distance $a_1$ of the 4 atom system, $R=2.16$~$\mathrm{\mu}$m. Although $\Delta E_{\mathrm{FS}}\ll V_{dd}$, fine-structure may be resolved in the Rydberg excitation process and hence is relevant for our problem.

In the following, we illustrate how it is nonetheless possible to obtain a simple effective state space and dipole-dipole coupling with negative sign as employed in the main article by applying an external magnetic field. Including spin, we denote the $l=0$ states with $\ket{s_{m_{s}}}$ and the $l=1$ states  with $\ket{p_{\mathcal{J}}}$, where the determination can either be done with the quantum numbers of the total angular momentum, $\mathcal{J}=j, m_j$ or the quantum numbers of the separate orbital $m_l$ and spin quantum numbers $m_{s_p}$, thus $\mathcal{J}=m_l, m_{s_p}$. Levels with different magnetic $m$-numbers typically experience different Zeeman shifts when applying an external magnetic field. We restrict ourselves to the $l=0,1$ states and use the two-atom bases $\mathcal{B_{J}}=\{\ket{s_{m_s}p_{\mathcal{J'}}},\ket{p_{\mathcal{J'}}s_{m_s}}\}_{m_s \in \{\downarrow,\uparrow\},\ J'\in \mathcal{K}}$, where $\mathcal{K}$ is the set of all possible quantum number realizations of the p-state. 
Restriction to these bases and shifting the zero point energy to $E_{s} + E_{p_{j=1/2}}$, the total Hamiltonian for two dipole-coupled atoms under the influence of an external magnetic field $B_{z}$ reads:
\begin{equation}
\label{Htot}
 \hat{H}_{\mathrm{tot}}=\hat{V}_{dd}(R) + \hat{H}_{\mathcal{SO}}+\hat{H}_{\mathcal{MF}}(B_z),
\end{equation}
where $\hat{V}_{dd}(R)$ is the dipole-dipole interaction with interatomic axis chosen orthogonal to the quantisation axis, $\hat{H}_{\mathcal{SO}}$ is the sum over the single atom spin-orbit operators and \begin{equation}
\hat{H}_{\mathcal{MF}}(B_z) = \mu_{\mathrm{B}}B_{z}\sum_{i=1}^{2}\bigl(\hat{L}_{z}^{(i)}+2\hat{S}_{z}^{(i)}\bigr)
\end{equation}
describes the interaction with a magnetic field oriented along the quantization axis. The latter is diagonal in $\mathcal{B}_{m_l,m_{s_p}}$ with matrix elements
\begin{align}
 &\bra{s_{m_s}p_{m_l,m_{s_p}}}\hat{H}_{\mathcal{MF}}(B_z)\ket{s_{m_s}p_{m_l,m_{s_p}}}/\mu_{\mathrm{B}}B_{z}
 \CR
 &=2(m_s + m_{s_{p}}) + m_l.
\end{align}
The spin-orbit Hamiltonian is diagonal in $\mathcal{B}_{j,m_j}$ with matrix elements
\begin{equation}
\bra{s_{m_s}p_{j,m_j}}\hat{H}_{\mathcal{SO}}\ket{s_{m_s}p_{j,m_j}}=\Delta E_{\mathrm{FS}}\delta_{j,3/2}.
\end{equation}
It turns out that the suitable basis of $\hat{H}_{\mathrm{tot}}$ is $\mathcal{B}_{j,m_j}$, thus we first write $\hat{V}_{dd}(R) + \hat{H}_{\mathcal{MF}}(B_z)$ in their natural basis $\mathcal{B}_{m_l,m_{s_p}}$ and perform an orthogonal transformation to $\mathcal{B}_{j,m_j}$.\newline
For a magnetic field strength of $B_z=240$~G, we find a subspace spanned by $\{\ket{s_{\uparrow}p_{j,m_j}},\ket{p_{j,m_j}s_{\uparrow}}\}_{j=m_j=3/2}$, that decouples from all the other states with a probability of 87.9\%.
The magnetic field shifts both states about $\Delta E_{\mathcal{MF}}=1008$~MHz. If we assume perfect decoupling and 
shift the zero of energy to $E_{p_{j=3/2}} + E_{s} + 3 \mu_{B} B_z $, we end up with the effective Hamiltonian
\begin{equation}
\hat{H}_{\mathrm{eff}}=
 \begin{pmatrix}
  0 & -\frac{d_{\mathrm{rad}}^2(44)}{6R^3}\\
-\frac{d_{\mathrm{rad}}^2(44)}{6R^3} & 0
 \end{pmatrix},
\label{effmatrix}
\end{equation}
with $d_{\mathrm{rad}}(44)=2495$~au.
This yields the parameter $\mu~=~d_{\mathrm{rad}}(\nu)/\sqrt{6}$, quoted in the main text. Imperfections of the decoupling cause slight 
modifications of functional form and strength of the off-diagonal couplings in \bref{effmatrix}, which are not used in the main article for simplicity.

We have however explicitly verified the state space reduction just described, neglecting spin-orbit coupling for tractable simulations.
To this end we have run simulations as shown in Fig.~2 of the main article, using 
an electronic basis $\ket{\pi_n,m_l}=\ket{s\dots (p,m_l)\dots s}$, see \cite{moebius:cradle}, with explicit Zeeman shifts $m_l \mu_B B_z$. We neglect (the small) spin-orbit coupling here to obtain a computationally more tractable problem. 
The reduced state space description in the main article is found adequate, residual quantitative differences that we find are deviations of the effective potential from a $R^{-3}$ form towards $R^{-4}$ at short distances, as well as modified exciton states very close to the conical intersection. Neither qualitatively affects motional and non-adiabatic dynamics, nor most importantly the described entanglement generation between position and exciton state. A more detailed study of the model involving the full spin degree of freedom without magnetic field will be subject of future work.

Importantly, matrix elements in \bref{effmatrix} are negative as in Eq.~2 of the main text. This is crucial to realize the trimer conical intersection between the upper two surfaces.

\sew{
Finally we show in \fref{fig:effective_energies} how the effective model described in the main article (utilizing only states $\ket{s}$ and $\ket{p}$ per atom) approximates full atomic interaction potentials obtained by exact diagonalisation of \eref{Htot} for $B_z=240 G$. We choose the example $\nu=80$ relevant for Fig.~3 in the main article. Each atomic basis includes states $\nu \in \{78 \cdots 82\}$, $l \in \{0 \cdots 3\}$ with all available $j$, $m_j$ states fulfilling $M=m_{j1} + m_{j2}=1$ or $M=2$, where $m_{jn}$ is the magnetic quantum number of atom $n$.

\begin{figure}[htb]
\centering
\epsfig{file={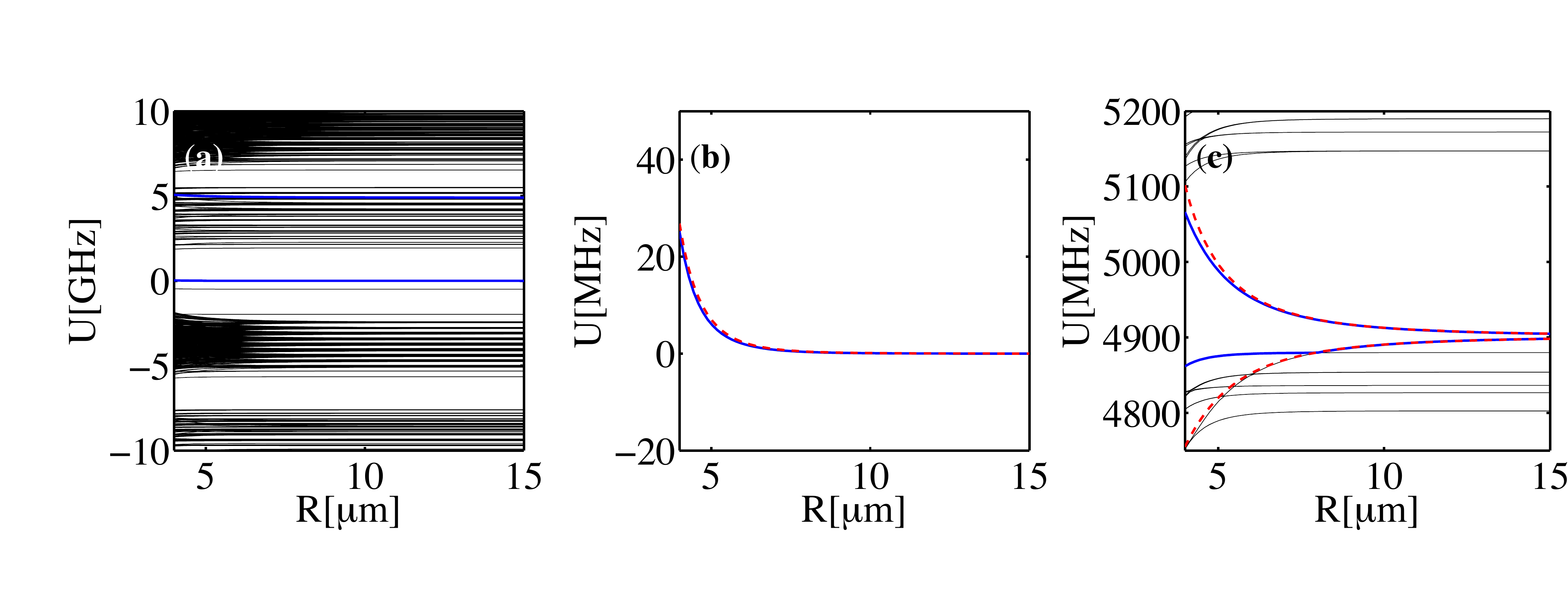},width=0.99\columnwidth} 
\caption{\label{fig:effective_energies}
Our effective model (equations 1-3  in the main article) in the context of the full space of molecular potentials obtained from exact diagonalisation. (a) Set of molecular potentials in the energetic vicinity of the $\ket{80s_{1/2}}\ket{80s_{1/2}}$-pair state (whose energy is set to zero). The two blue lines are the potentials of interest, shown more detailed in panels (b-c). (b) Zoom onto (blue) $\ket{80s}\ket{80s}$ pair potential from exact diagonalisation, (red-dashed) model (3).  (c) Zoom onto (blue) $\ket{80s_{1/2}}\ket{80p_{3/2}}\pm \ket{80p_{3/2}}\ket{80s_{1/2}}$ pair potentials from exact diagonalisation, (red-dashed) model (2,3). 
}
\end{figure}
Parameters for the model in equations (2,3) of the main article are fitted in the red-dashed lines, we obtain $C_{6,ss}= - 7.6\times 10^{20}$ au.~and $\mu = 3374$ au. It is seen in panel (c) that for $R>5 {\mu}m$ the relevant repulsive potential is energetically well separated from other energy surfaces, justifying our reduction of the state space. The nearest neighbouring pair 
states also visible in panel (c) belong to the $\ket{81d,f} \ket{79s}$, $\ket{81s} \ket{79d,f}$, $\ket{80d,f} \ket{80s}$ finestructure manifolds.
}


\begin{thebibliography}{37}
\expandafter\ifx\csname natexlab\endcsname\relax\def\natexlab#1{#1}\fi
\expandafter\ifx\csname bibnamefont\endcsname\relax
  \def\bibnamefont#1{#1}\fi
\expandafter\ifx\csname bibfnamefont\endcsname\relax
  \def\bibfnamefont#1{#1}\fi
\expandafter\ifx\csname citenamefont\endcsname\relax
  \def\citenamefont#1{#1}\fi
\expandafter\ifx\csname url\endcsname\relax
  \def\url#1{\texttt{#1}}\fi
\expandafter\ifx\csname urlprefix\endcsname\relax\def\urlprefix{URL }\fi
\providecommand{\bibinfo}[2]{#2}
\providecommand{\eprint}[2][]{\url{#2}}

\bibitem[{\citenamefont{Frenkel}(1931)}]{frenkel_exciton}
\bibinfo{author}{\bibfnamefont{J.}~\bibnamefont{Frenkel}},
  \bibinfo{journal}{Phys. Rev.} \textbf{\bibinfo{volume}{37}},
  \bibinfo{pages}{17} (\bibinfo{year}{1931}).

\bibitem[{\citenamefont{May and K{\"u}hn}(2001)}]{book:maykuehn}
\bibinfo{author}{\bibfnamefont{V.}~\bibnamefont{May}} \bibnamefont{and}
  \bibinfo{author}{\bibfnamefont{O.}~\bibnamefont{K{\"u}hn}},
  \emph{\bibinfo{title}{Charge and Energy Transfer Dynamics in Molecular
  Systems}} (\bibinfo{publisher}{Wiley-VCH, Berlin}, \bibinfo{year}{2001}).

\bibitem[{\citenamefont{Park et~al.}(2011{\natexlab{a}})\citenamefont{Park,
  Tanner, Claessens, Shuman, and Gallagher}}]{park:dipdipbroadening}
\bibinfo{author}{\bibfnamefont{H.}~\bibnamefont{Park}},
  \bibinfo{author}{\bibfnamefont{P.~J.} \bibnamefont{Tanner}},
  \bibinfo{author}{\bibfnamefont{B.~J.} \bibnamefont{Claessens}},
  \bibinfo{author}{\bibfnamefont{E.~S.} \bibnamefont{Shuman}},
  \bibnamefont{and} \bibinfo{author}{\bibfnamefont{T.~F.}
  \bibnamefont{Gallagher}}, \bibinfo{journal}{Phys. Rev. A}
  \textbf{\bibinfo{volume}{84}}, \bibinfo{pages}{022704}
  (\bibinfo{year}{2011}{\natexlab{a}}).

\bibitem[{\citenamefont{Park et~al.}(2011{\natexlab{b}})\citenamefont{Park,
  Shuman, and Gallagher}}]{park:dipdipionization}
\bibinfo{author}{\bibfnamefont{H.}~\bibnamefont{Park}},
  \bibinfo{author}{\bibfnamefont{E.~S.} \bibnamefont{Shuman}},
  \bibnamefont{and} \bibinfo{author}{\bibfnamefont{T.~F.}
  \bibnamefont{Gallagher}}, \bibinfo{journal}{Phys. Rev. A}
  \textbf{\bibinfo{volume}{84}}, \bibinfo{pages}{052708}
  (\bibinfo{year}{2011}{\natexlab{b}}).

\bibitem[{\citenamefont{Li et~al.}(2005)\citenamefont{Li, Tanner, and
  Gallagher}}]{li_gallagher:dipdipexcit}
\bibinfo{author}{\bibfnamefont{W.}~\bibnamefont{Li}},
  \bibinfo{author}{\bibfnamefont{P.~J.} \bibnamefont{Tanner}},
  \bibnamefont{and} \bibinfo{author}{\bibfnamefont{T.~F.}
  \bibnamefont{Gallagher}}, \bibinfo{journal}{Phys. Rev. Lett.}
  \textbf{\bibinfo{volume}{94}}, \bibinfo{pages}{173001}
  (\bibinfo{year}{2005}).

\bibitem[{\citenamefont{Westermann et~al.}(2006)\citenamefont{Westermann,
  Amthor, de~Oliveira, Deiglmayr, {Reetz-Lamour}, and
  Weidem{\"u}ller}}]{westermann:transfer}
\bibinfo{author}{\bibfnamefont{S.}~\bibnamefont{Westermann}},
  \bibinfo{author}{\bibfnamefont{T.}~\bibnamefont{Amthor}},
  \bibinfo{author}{\bibfnamefont{A.}~\bibnamefont{de~Oliveira}},
  \bibinfo{author}{\bibfnamefont{J.}~\bibnamefont{Deiglmayr}},
  \bibinfo{author}{\bibfnamefont{M.}~\bibnamefont{{Reetz-Lamour}}},
  \bibnamefont{and}
  \bibinfo{author}{\bibfnamefont{M.}~\bibnamefont{Weidem{\"u}ller}},
  \bibinfo{journal}{Eur. Phys. J. D} \textbf{\bibinfo{volume}{40}},
  \bibinfo{pages}{37} (\bibinfo{year}{2006}).

\bibitem[{\citenamefont{M{\"u}lken et~al.}(2007)\citenamefont{M{\"u}lken,
  Blumen, Amthor, Giese, Reetz-Lamour, and
  Weidem\"uller}}]{muelken:excitontransfer}
\bibinfo{author}{\bibfnamefont{O.}~\bibnamefont{M{\"u}lken}},
  \bibinfo{author}{\bibfnamefont{A.}~\bibnamefont{Blumen}},
  \bibinfo{author}{\bibfnamefont{T.}~\bibnamefont{Amthor}},
  \bibinfo{author}{\bibfnamefont{C.}~\bibnamefont{Giese}},
  \bibinfo{author}{\bibfnamefont{M.}~\bibnamefont{Reetz-Lamour}},
  \bibnamefont{and}
  \bibinfo{author}{\bibfnamefont{M.}~\bibnamefont{Weidem\"uller}},
  \bibinfo{journal}{Phys. Rev. Lett.} \textbf{\bibinfo{volume}{99}},
  \bibinfo{pages}{090601} (\bibinfo{year}{2007}).

\bibitem[{\citenamefont{Bettelli et~al.}(2013)\citenamefont{Bettelli, Maxwell,
  Fernholz, Adams, Lesanovsky, and Ates}}]{bettelli:emerglatt}
\bibinfo{author}{\bibfnamefont{S.}~\bibnamefont{Bettelli}},
  \bibinfo{author}{\bibfnamefont{D.}~\bibnamefont{Maxwell}},
  \bibinfo{author}{\bibfnamefont{T.}~\bibnamefont{Fernholz}},
  \bibinfo{author}{\bibfnamefont{C.~S.} \bibnamefont{Adams}},
  \bibinfo{author}{\bibfnamefont{I.}~\bibnamefont{Lesanovsky}},
  \bibnamefont{and} \bibinfo{author}{\bibfnamefont{C.}~\bibnamefont{Ates}},
  \bibinfo{journal}{Phys. Rev. A} \textbf{\bibinfo{volume}{88}},
  \bibinfo{pages}{043436} (\bibinfo{year}{2013}).

\bibitem[{\citenamefont{G{\"u}nter et~al.}(2013)\citenamefont{G{\"u}nter,
  Schempp, {Robert-de-Saint-Vincent}, Gavryusev, Helmrich, Hofmann, Whitlock,
  and Weidem{\"u}ller}}]{guenter:observingtransp}
\bibinfo{author}{\bibfnamefont{G.}~\bibnamefont{G{\"u}nter}},
  \bibinfo{author}{\bibfnamefont{H.}~\bibnamefont{Schempp}},
  \bibinfo{author}{\bibfnamefont{M.}~\bibnamefont{{Robert-de-Saint-Vincent}}},
  \bibinfo{author}{\bibfnamefont{V.}~\bibnamefont{Gavryusev}},
  \bibinfo{author}{\bibfnamefont{S.}~\bibnamefont{Helmrich}},
  \bibinfo{author}{\bibfnamefont{C.~S.} \bibnamefont{Hofmann}},
  \bibinfo{author}{\bibfnamefont{S.}~\bibnamefont{Whitlock}}, \bibnamefont{and}
  \bibinfo{author}{\bibfnamefont{M.}~\bibnamefont{Weidem{\"u}ller}},
  \bibinfo{journal}{Science} \textbf{\bibinfo{volume}{342}},
  \bibinfo{pages}{954} (\bibinfo{year}{2013}).

\bibitem[{\citenamefont{Ravets et~al.}(2014)\citenamefont{Ravets, Labuhn,
  Barredo, B{\'e}guin, Lahaye, and Browaeys}}]{ravets:foersterdipdip}
\bibinfo{author}{\bibfnamefont{S.}~\bibnamefont{Ravets}},
  \bibinfo{author}{\bibfnamefont{H.}~\bibnamefont{Labuhn}},
  \bibinfo{author}{\bibfnamefont{D.}~\bibnamefont{Barredo}},
  \bibinfo{author}{\bibfnamefont{L.}~\bibnamefont{B{\'e}guin}},
  \bibinfo{author}{\bibfnamefont{T.}~\bibnamefont{Lahaye}}, \bibnamefont{and}
  \bibinfo{author}{\bibfnamefont{A.}~\bibnamefont{Browaeys}}
  (\bibinfo{year}{2014}), \bibinfo{note}{arXiv:1405.7804}.

\bibitem[{\citenamefont{Barredo et~al.}(2014)\citenamefont{Barredo, Labuhn,
  Ravets, Lahaye, Browaeys, and Adams}}]{barredo:threespinchain}
\bibinfo{author}{\bibfnamefont{D.}~\bibnamefont{Barredo}},
  \bibinfo{author}{\bibfnamefont{H.}~\bibnamefont{Labuhn}},
  \bibinfo{author}{\bibfnamefont{S.}~\bibnamefont{Ravets}},
  \bibinfo{author}{\bibfnamefont{T.}~\bibnamefont{Lahaye}},
  \bibinfo{author}{\bibfnamefont{A.}~\bibnamefont{Browaeys}}, \bibnamefont{and}
  \bibinfo{author}{\bibfnamefont{C.~S.} \bibnamefont{Adams}}
  (\bibinfo{year}{2014}), \bibinfo{note}{arXiv:1408.1055}.

\bibitem[{\citenamefont{Gallagher}(1994)}]{book:gallagher}
\bibinfo{author}{\bibfnamefont{T.~F.} \bibnamefont{Gallagher}},
  \emph{\bibinfo{title}{Rydberg Atoms}} (\bibinfo{publisher}{Cambridge
  University Press}, \bibinfo{year}{1994}).

\bibitem[{\citenamefont{Beterov et~al.}(2009)\citenamefont{Beterov, Tretyakov,
  Ryabtsev, Entin, Ekers, and
  Bezuglov}}]{Beterov-Bezuglov-IonizationofRydberg-2009}
\bibinfo{author}{\bibfnamefont{I.~I.} \bibnamefont{Beterov}},
  \bibinfo{author}{\bibfnamefont{D.~B.} \bibnamefont{Tretyakov}},
  \bibinfo{author}{\bibfnamefont{I.~I.} \bibnamefont{Ryabtsev}},
  \bibinfo{author}{\bibfnamefont{V.~M.} \bibnamefont{Entin}},
  \bibinfo{author}{\bibfnamefont{A.}~\bibnamefont{Ekers}}, \bibnamefont{and}
  \bibinfo{author}{\bibfnamefont{N.~N.} \bibnamefont{Bezuglov}},
  \bibinfo{journal}{New J. Phys.} \textbf{\bibinfo{volume}{11}},
  \bibinfo{pages}{013052} (\bibinfo{year}{2009}).

\bibitem[{\citenamefont{Ates et~al.}(2008)\citenamefont{Ates, Eisfeld, and
  Rost}}]{cenap:motion}
\bibinfo{author}{\bibfnamefont{C.}~\bibnamefont{Ates}},
  \bibinfo{author}{\bibfnamefont{A.}~\bibnamefont{Eisfeld}}, \bibnamefont{and}
  \bibinfo{author}{\bibfnamefont{J.~M.} \bibnamefont{Rost}},
  \bibinfo{journal}{New J. Phys.} \textbf{\bibinfo{volume}{10}},
  \bibinfo{pages}{045030} (\bibinfo{year}{2008}).

\bibitem[{\citenamefont{W{\"u}ster et~al.}(2010)\citenamefont{W{\"u}ster, Ates,
  Eisfeld, and Rost}}]{wuester:cradle}
\bibinfo{author}{\bibfnamefont{S.}~\bibnamefont{W{\"u}ster}},
  \bibinfo{author}{\bibfnamefont{C.}~\bibnamefont{Ates}},
  \bibinfo{author}{\bibfnamefont{A.}~\bibnamefont{Eisfeld}}, \bibnamefont{and}
  \bibinfo{author}{\bibfnamefont{J.~M.} \bibnamefont{Rost}},
  \bibinfo{journal}{Phys. Rev. Lett.} \textbf{\bibinfo{volume}{105}},
  \bibinfo{pages}{053004} (\bibinfo{year}{2010}).

\bibitem[{\citenamefont{M{\"o}bius et~al.}(2011)\citenamefont{M{\"o}bius,
  W{\"u}ster, Ates, Eisfeld, and Rost}}]{moebius:cradle}
\bibinfo{author}{\bibfnamefont{S.}~\bibnamefont{M{\"o}bius}},
  \bibinfo{author}{\bibfnamefont{S.}~\bibnamefont{W{\"u}ster}},
  \bibinfo{author}{\bibfnamefont{C.}~\bibnamefont{Ates}},
  \bibinfo{author}{\bibfnamefont{A.}~\bibnamefont{Eisfeld}}, \bibnamefont{and}
  \bibinfo{author}{\bibfnamefont{J.~M.} \bibnamefont{Rost}},
  \bibinfo{journal}{J. Phys. B: At. Mol. Opt. Phys.}
  \textbf{\bibinfo{volume}{44}}, \bibinfo{pages}{184011}
  (\bibinfo{year}{2011}).

\bibitem[{\citenamefont{Asadian et~al.}(2010)\citenamefont{Asadian, Tiersch,
  Guerreschi, Cai, Popescu, and Briegel}}]{asadian:motion}
\bibinfo{author}{\bibfnamefont{A.}~\bibnamefont{Asadian}},
  \bibinfo{author}{\bibfnamefont{M.}~\bibnamefont{Tiersch}},
  \bibinfo{author}{\bibfnamefont{G.~G.} \bibnamefont{Guerreschi}},
  \bibinfo{author}{\bibfnamefont{J.}~\bibnamefont{Cai}},
  \bibinfo{author}{\bibfnamefont{S.}~\bibnamefont{Popescu}}, \bibnamefont{and}
  \bibinfo{author}{\bibfnamefont{H.~J.} \bibnamefont{Briegel}},
  \bibinfo{journal}{New J. Phys.} \textbf{\bibinfo{volume}{12}},
  \bibinfo{pages}{075019} (\bibinfo{year}{2010}).

\bibitem[{\citenamefont{Eisfeld}(2011)}]{alex:heart}
\bibinfo{author}{\bibfnamefont{A.}~\bibnamefont{Eisfeld}},
  \bibinfo{journal}{Chemical Physics} \textbf{\bibinfo{volume}{379}},
  \bibinfo{pages}{33 } (\bibinfo{year}{2011}).

\bibitem[{\citenamefont{Saikin et~al.}(2013)\citenamefont{Saikin, Eisfeld,
  Valleau, and {Aspuru-Guzik}}}]{saikin:excitonreview}
\bibinfo{author}{\bibfnamefont{S.~K.} \bibnamefont{Saikin}},
  \bibinfo{author}{\bibfnamefont{A.}~\bibnamefont{Eisfeld}},
  \bibinfo{author}{\bibfnamefont{S.}~\bibnamefont{Valleau}}, \bibnamefont{and}
  \bibinfo{author}{\bibfnamefont{A.}~\bibnamefont{{Aspuru-Guzik}}},
  \bibinfo{journal}{Nanophotonics} \textbf{\bibinfo{volume}{2}},
  \bibinfo{pages}{21} (\bibinfo{year}{2013}).

\bibitem[{\citenamefont{K{\"u}hn and Lochbrunner}(2011)}]{kuehn:excitonreview}
\bibinfo{author}{\bibfnamefont{O.}~\bibnamefont{K{\"u}hn}} \bibnamefont{and}
  \bibinfo{author}{\bibfnamefont{S.}~\bibnamefont{Lochbrunner}},
  \bibinfo{journal}{Semiconductors and Semimetals}
  \textbf{\bibinfo{volume}{85}}, \bibinfo{pages}{47} (\bibinfo{year}{2011}).

\bibitem[{\citenamefont{W{\"u}ster et~al.}(2011)\citenamefont{W{\"u}ster,
  Eisfeld, and Rost}}]{wuester:CI}
\bibinfo{author}{\bibfnamefont{S.}~\bibnamefont{W{\"u}ster}},
  \bibinfo{author}{\bibfnamefont{A.}~\bibnamefont{Eisfeld}}, \bibnamefont{and}
  \bibinfo{author}{\bibfnamefont{J.~M.} \bibnamefont{Rost}},
  \bibinfo{journal}{Phys. Rev. Lett.} \textbf{\bibinfo{volume}{106}},
  \bibinfo{pages}{153002} (\bibinfo{year}{2011}).

\bibitem[{\citenamefont{Domcke et~al.}(2004)\citenamefont{Domcke, Yarkony, and
  K{\"o}ppel}}]{domke:yarkony:koeppel:CIs}
\bibinfo{author}{\bibfnamefont{W.}~\bibnamefont{Domcke}},
  \bibinfo{author}{\bibfnamefont{D.~R.} \bibnamefont{Yarkony}},
  \bibnamefont{and}
  \bibinfo{author}{\bibfnamefont{H.}~\bibnamefont{K{\"o}ppel}},
  \emph{\bibinfo{title}{Conical Intersections}} (\bibinfo{publisher}{World
  Scientific}, \bibinfo{year}{2004}).

\bibitem[{\citenamefont{Li et~al.}(2013)\citenamefont{Li, Dudin, and
  Kuzmich}}]{li:lightatomentangle}
\bibinfo{author}{\bibfnamefont{L.}~\bibnamefont{Li}},
  \bibinfo{author}{\bibfnamefont{Y.~O.} \bibnamefont{Dudin}}, \bibnamefont{and}
  \bibinfo{author}{\bibfnamefont{A.}~\bibnamefont{Kuzmich}},
  \bibinfo{journal}{Nature} \textbf{\bibinfo{volume}{498}},
  \bibinfo{pages}{466} (\bibinfo{year}{2013}).

\bibitem[{\citenamefont{Mukherjee et~al.}(2011)\citenamefont{Mukherjee, Millen,
  Nath, Jones, and Pohl}}]{rick:Rydberglattice}
\bibinfo{author}{\bibfnamefont{R.}~\bibnamefont{Mukherjee}},
  \bibinfo{author}{\bibfnamefont{J.}~\bibnamefont{Millen}},
  \bibinfo{author}{\bibfnamefont{R.}~\bibnamefont{Nath}},
  \bibinfo{author}{\bibfnamefont{M.~P.~A.} \bibnamefont{Jones}},
  \bibnamefont{and} \bibinfo{author}{\bibfnamefont{T.}~\bibnamefont{Pohl}},
  \bibinfo{journal}{J. Phys. B: At. Mol. Opt. Phys.}
  \textbf{\bibinfo{volume}{33}}, \bibinfo{pages}{184010}
  (\bibinfo{year}{2011}).

\bibitem[{sup()}]{sup:info}
\bibinfo{note}{See Supplemental Material for additional details regarding the employed quantum-classical algorithm, the Rydberg trimer subunit, our purity and entanglement measure, extraction of total atomic densities and the realization of isotropic dipole-dipole interactions, which includes Refs.~\cite{tucker73_geometry_ci,mead79_bo_with_ci,hill:wootters:qbits,wootters:mixed,haroche:li_finesplitting}}.

\bibitem[{\citenamefont{Carrington}(1974)}]{tucker73_geometry_ci}
\bibinfo{author}{\bibfnamefont{T.}~\bibnamefont{Carrington}},
  \bibinfo{journal}{Accounts of Chemical Research}
  \textbf{\bibinfo{volume}{7}}, \bibinfo{pages}{20} (\bibinfo{year}{1974}).

\bibitem[{\citenamefont{Mead and Truhlar}(1979)}]{mead79_bo_with_ci}
\bibinfo{author}{\bibfnamefont{C.~A.} \bibnamefont{Mead}} \bibnamefont{and}
  \bibinfo{author}{\bibfnamefont{D.~G.} \bibnamefont{Truhlar}},
  \bibinfo{journal}{The Journal of Chemical Physics}
  \textbf{\bibinfo{volume}{70}}, \bibinfo{pages}{2284} (\bibinfo{year}{1979}).

\bibitem[{\citenamefont{Hill and Wootters}(1997)}]{hill:wootters:qbits}
\bibinfo{author}{\bibfnamefont{S.}~\bibnamefont{Hill}} \bibnamefont{and}
  \bibinfo{author}{\bibfnamefont{W.~K.} \bibnamefont{Wootters}},
  \bibinfo{journal}{Phys. Rev. Lett.} \textbf{\bibinfo{volume}{78}},
  \bibinfo{pages}{5022} (\bibinfo{year}{1997}).

\bibitem[{\citenamefont{Wootters}(1998)}]{wootters:mixed}
\bibinfo{author}{\bibfnamefont{W.~K.} \bibnamefont{Wootters}},
  \bibinfo{journal}{Phys. Rev. Lett.} \textbf{\bibinfo{volume}{80}},
  \bibinfo{pages}{2245} (\bibinfo{year}{1998}).

\bibitem[{\citenamefont{Goy et~al.}(1986)\citenamefont{Goy, Liang, and
  Haroche}}]{haroche:li_finesplitting}
\bibinfo{author}{\bibfnamefont{P.}~\bibnamefont{Goy}},
  \bibinfo{author}{\bibfnamefont{J.}~\bibnamefont{Liang}},
\bibinfo{author}{\bibfnamefont{M.}~\bibnamefont{Gross}}, \bibnamefont{and}
  \bibinfo{author}{\bibfnamefont{S.}~\bibnamefont{Haroche}},
  \bibinfo{journal}{Phys. Rev. A} \textbf{\bibinfo{volume}{34}},
  \bibinfo{pages}{2889} (\bibinfo{year}{1986}).


\bibitem[{\citenamefont{Tully and Preston}(1971)}]{tully:hopping2}
\bibinfo{author}{\bibfnamefont{J.~C.} \bibnamefont{Tully}} \bibnamefont{and}
  \bibinfo{author}{\bibfnamefont{R.~K.} \bibnamefont{Preston}},
  \bibinfo{journal}{J. Chem. Phys.} \textbf{\bibinfo{volume}{55}},
  \bibinfo{pages}{562} (\bibinfo{year}{1971}).

\bibitem[{\citenamefont{{Hammes-Schiffer} and
  Tully}(1994)}]{tully:hopping:veloadjust}
\bibinfo{author}{\bibfnamefont{S.}~\bibnamefont{{Hammes-Schiffer}}}
  \bibnamefont{and} \bibinfo{author}{\bibfnamefont{J.~C.} \bibnamefont{Tully}},
  \bibinfo{journal}{J. Chem. Phys.} \textbf{\bibinfo{volume}{101}},
  \bibinfo{pages}{4657} (\bibinfo{year}{1994}).

\bibitem[{\citenamefont{Barbatti}(2011)}]{barbatti:review_tully}
\bibinfo{author}{\bibfnamefont{M.}~\bibnamefont{Barbatti}},
  \bibinfo{journal}{Wiley Interdisciplinary Reviews-Computational Molecular
  Science} \textbf{\bibinfo{volume}{1}}, \bibinfo{pages}{620}
  (\bibinfo{year}{2011}).

\bibitem[{\citenamefont{Jasper and Truhlar}({2005})}]{truhlar:tully_CI_acc}
\bibinfo{author}{\bibfnamefont{A.}~\bibnamefont{Jasper}} \bibnamefont{and}
  \bibinfo{author}{\bibfnamefont{D.}~\bibnamefont{Truhlar}},
  \bibinfo{journal}{J. Chem. Phys.} \textbf{\bibinfo{volume}{{122}}}
  (\bibinfo{year}{{2005}}), ISSN \bibinfo{issn}{{0021-9606}}.

\bibitem[{\citenamefont{M{\"o}bius et~al.}(2013)\citenamefont{M{\"o}bius,
  Genkin, Eisfeld, Wuster, and Rost}}]{moebius:bobbels}
\bibinfo{author}{\bibfnamefont{S.}~\bibnamefont{M{\"o}bius}},
  \bibinfo{author}{\bibfnamefont{M.}~\bibnamefont{Genkin}},
  \bibinfo{author}{\bibfnamefont{A.}~\bibnamefont{Eisfeld}},
  \bibinfo{author}{\bibfnamefont{S.}~\bibnamefont{Wuster}}, \bibnamefont{and}
  \bibinfo{author}{\bibfnamefont{J.~M.} \bibnamefont{Rost}},
  \bibinfo{journal}{Phys. Rev. A} \textbf{\bibinfo{volume}{87}},
  \bibinfo{pages}{051602} (\bibinfo{year}{2013}).

\bibitem[{\citenamefont{Dennis et~al.}(2013)\citenamefont{Dennis, Hope, and
  Johnsson}}]{xmds}
\bibinfo{author}{\bibfnamefont{G.~R.} \bibnamefont{Dennis}},
  \bibinfo{author}{\bibfnamefont{J.~J.} \bibnamefont{Hope}}, \bibnamefont{and}
  \bibinfo{author}{\bibfnamefont{M.~T.} \bibnamefont{Johnsson}},
  \bibinfo{journal}{Comput. Phys. Commun.} \textbf{\bibinfo{volume}{184}},
  \bibinfo{pages}{201} (\bibinfo{year}{2013}).

\bibitem[{foo()}]{footnote:notrimer}
\bibinfo{note}{Considering a trimer subunit only makes sense in a moment where
  three atoms have clearly closer mutual separations than all others.}

\bibitem[{\citenamefont{Teller}(1937)}]{teller1937crossing}
\bibinfo{author}{\bibfnamefont{E.}~\bibnamefont{Teller}},
  \bibinfo{journal}{J.Phys. Chem.} \textbf{\bibinfo{volume}{41}},
  \bibinfo{pages}{109} (\bibinfo{year}{1937}).

\bibitem[{\citenamefont{McConnell et~al.}(2010)\citenamefont{McConnell, Li, and
  Brudvig}}]{mcconell:artificialLHCreview}
\bibinfo{author}{\bibfnamefont{I.}~\bibnamefont{McConnell}},
  \bibinfo{author}{\bibfnamefont{G.}~\bibnamefont{Li}}, \bibnamefont{and}
  \bibinfo{author}{\bibfnamefont{G.~W.} \bibnamefont{Brudvig}},
  \bibinfo{journal}{Chemistry {\&} Biology} \textbf{\bibinfo{volume}{17}},
  \bibinfo{pages}{434} (\bibinfo{year}{2010}).

\bibitem[{\citenamefont{Schwarzkopf et~al.}(2011)\citenamefont{Schwarzkopf,
  Sapiro, and Raithel}}]{schwarzkopf:correlations}
\bibinfo{author}{\bibfnamefont{A.}~\bibnamefont{Schwarzkopf}},
  \bibinfo{author}{\bibfnamefont{R.~E.} \bibnamefont{Sapiro}},
  \bibnamefont{and} \bibinfo{author}{\bibfnamefont{G.}~\bibnamefont{Raithel}},
  \bibinfo{journal}{Phys. Rev. Lett.} \textbf{\bibinfo{volume}{107}},
  \bibinfo{pages}{103001} (\bibinfo{year}{2011}).

\bibitem[{\citenamefont{Olmos et~al.}(2011)\citenamefont{Olmos, Li,
  Hofferberth, and Lesanovsky}}]{olmos:amplification}
\bibinfo{author}{\bibfnamefont{B.}~\bibnamefont{Olmos}},
  \bibinfo{author}{\bibfnamefont{W.}~\bibnamefont{Li}},
  \bibinfo{author}{\bibfnamefont{S.}~\bibnamefont{Hofferberth}},
  \bibnamefont{and}
  \bibinfo{author}{\bibfnamefont{I.}~\bibnamefont{Lesanovsky}},
  \bibinfo{journal}{Phys. Rev. A} \textbf{\bibinfo{volume}{84}},
  \bibinfo{pages}{041607(R)} (\bibinfo{year}{2011}).

\bibitem[{\citenamefont{G{\"u}nter et~al.}(2012)\citenamefont{G{\"u}nter,
  de~Saint-Vincent, Schempp, Hofmann, Whitlock, and
  Weidem{\"u}ller}}]{guenter:EIT}
\bibinfo{author}{\bibfnamefont{G.}~\bibnamefont{G{\"u}nter}},
  \bibinfo{author}{\bibfnamefont{M.~R.} \bibnamefont{de~Saint-Vincent}},
  \bibinfo{author}{\bibfnamefont{H.}~\bibnamefont{Schempp}},
  \bibinfo{author}{\bibfnamefont{C.~S.} \bibnamefont{Hofmann}},
  \bibinfo{author}{\bibfnamefont{S.}~\bibnamefont{Whitlock}}, \bibnamefont{and}
  \bibinfo{author}{\bibfnamefont{M.}~\bibnamefont{Weidem{\"u}ller}},
  \bibinfo{journal}{Phys. Rev. Lett.} \textbf{\bibinfo{volume}{108}},
  \bibinfo{pages}{013002} (\bibinfo{year}{2012}).



\end{thebibliography}

\end{document}